\documentclass[useAMS,usegraphicx]{mn2e}
\usepackage{psfig}

\title[Optical Variability of Quasars]
{Intra-night Optical Variability of BL Lacs, Radio-Quiet Quasars and Radio-Loud Quasars}
\author[C.~S.\ Stalin et al.]
       {C.\ S.\ Stalin$^{1,2}$, Alok C.\ Gupta$^3$, Gopal-Krishna$^{4,5}$, 
Paul J.\ Wiita$^6$, and Ram Sagar$^{2,7}$\\
$^{1}$ IUCAA, Post Bag 4, Ganeshkhind, Pune 411 007, India \\
$^{2}$ State Observatory, Manora Peak, Naini Tal 263 129, India \\
$^{3}$ Harish-Chandra Research Institute, Chhatnag Road, Jhunsi, Allahabad 211019, India\\
$^{4}$ National Centre for Radio Astrophysics, TIFR, Pune University Campus, Post Bag 3, Pune 411
007, India\\
$^{5}$ Institut d'Astrophysique de Paris, 98 bis, Bd. Arago - 75014, Paris, France \\
$^{6}$ Department of Physics \& Astronomy, Georgia State University, P.O.\ Box 4106, Atlanta, 
Georgia 30302-4106, USA\\
$^{7}$ Aryabhatta Research Institute of Observational Sciences, Manora Peak, Nainital 263 129, India}

\date{Accepted 2004 xxxxx. Received 2004 xxxxx
           }

\pagerange{\pageref{firstpage}--\pageref{lastpage}}
\pubyear{2004}

\begin{document}

\maketitle

\label{firstpage}

\begin{abstract}
We report monitoring observations of 20 high-luminosity AGN, 12 of which are 
radio-quiet quasars (RQQs).  Intra-night optical variability (INOV) was
detected for
13 of the 20 objects, including 5 RQQs.  The
variations are distinctly stronger and more frequent for blazars than for the other AGN classes.
By combining these data with results obtained earlier in our
program, we have formed an enlarged sample consisting of 9 BL Lacs, 
19 RQQs and 11 lobe-dominated radio-loud quasars. The moderate level
of rapid optical variability found for both RQQs and radio lobe-dominated quasars 
argues against a direct link between INOV and radio-loudness. We 
supplemented the present observations of 3 BL Lacs with additional data from 
the literature.
In this extended sample of 12 well observed BL Lacs, stronger
INOV is found for the {\it EGRET} detected subset.
\end{abstract} 


\section{Introduction}

In a series of papers since 2003, we have reported results of a
program to search for intra-night optical variability (INOV; 
see Wagner \& Witzel 1995 for a review),
often called microvariability, in a sample of 26 optically luminous
active galactic nuclei (AGN), using the 104-cm Zeiss telescope
of the Aryabhatta Research Institute of Observational 
Sciences (ARIES), Naini Tal, India (Gopal-Krishna et al.\
 2003, GK03; Stalin et al.\ 2004a, St04a; Sagar et al.\
2004, Sa04; Stalin et al.\ 2004b, St04b; see also Stalin 2003).
These objects belong to the four major classes of luminous AGN,
namely, radio-quiet QSOs (RQQs), radio-loud lobe-dominated quasars
(LDQs), radio-loud core-dominated quasars (CDQs), and BL Lac objects (BLs).
The sample selection was such that the four classes are reasonably
well matched in the $z$--$M_B$ plane, with $z$ ranging from
0.17 to 2.2 and $M_B$ ranging from $-24.3$ to $-30.0$
(taking $H_0 = 50$ km s$^{-1}$ Mpc$^{-1}$ and $q_0 = 0$).

\begin{table}
\caption{The sample of 20 optically luminous AGN monitored in the present work}
\begin{tabular}{lllllrr} \hline
  &       &        &         &          &              &       \\
IAU Name    &  Type & ~~$B$    & ~$M_B$  & ~~ $z$   & $P_{opt}$$^\dag$ & log $R^{\ast}$$^{\ddag}$\\
            &       &        &         &          &    (\%)~~          &       \\
            &       &        &         &          &              &       \\ \hline
            &       &        &         &          &              &       \\
0003$+$158  &  LDQ  & 16.51  & $-$25.7 &  0.450   &  0.65  &     2.6  \\
0025$+$307  &  LDQ  & 15.79  & $-$26.7 &  0.500   &  ---   &     1.8  \\
0043$+$039  &  RQQ  & 16.00  & $-$26.0 &  0.385   &  0.27  &  $-$0.7  \\
0806$+$315  &  BL   & 15.70  & $-$25.0 &  0.220   &  ---   &     1.7  \\
0824$+$098  &  RQQ  & 15.50  & $-$25.6 &  0.260   &  ---   &     0.5 \\
0832$+$251  &  RQQ  & 16.10  & $-$25.5 &  0.331   &  ---   &     0.1 \\
0846$+$513  &  CDQ  & 16.28  & $-$29.4 &  1.860   &  ---   &     2.2  \\
0850$+$440  &  RQQ  & 16.40  & $-$26.1 &  0.513   &  ---   & $<-$0.5 \\
0931$+$437  &  RQQ  & 16.47  & $-$25.8 &  0.456   &  ---   &     0.1 \\
0935$+$416  &  RQQ  & 16.31  & $-$29.6 &  1.966   &  ---   & $<-$0.7 \\
0945$+$438  &  RQQ  & 16.28  & $-$24.5 &  0.226   &  ---   & $<-$0.1  \\
1029$+$329  &  RQQ  & 16.00  & $-$26.7 &  0.560   &  ---   & $<-$0.6  \\
1418$+$546  &  BL   & 16.17  & $-$23.7 &  0.152   &  7.5   &     3.1  \\
1422$+$424  &  RQQ  & 16.42  & $-$25.1 &  0.316   &  ---   & $<-$0.4  \\
1425$+$267  &  LDQ  & 15.78  & $-$26.0 &  0.366   &  1.9   &     2.0  \\
1444$+$407  &  RQQ  & 15.45  & $-$25.7 &  0.267   &  0.4   & $-$1.1  \\
1522$+$101  &  RQQ  & 16.20  & $-$28.4 &  1.324   &  0.3   & $<-$0.7  \\
1553$+$113  &  BL   & 15.00  & $-$26.8 &  0.360   &  ---   &     2.2  \\
1631$+$395  &  LDQ  & 16.48  & $-$27.8 &  1.023   &  1.1   &     1.6  \\
1750$+$507  &  RQQ  & 15.80  & $-$25.6 &  0.300   &  ---   &     0.7  \\
            &       &        &         &          &        &       \\  \hline
\end{tabular}

\hspace*{-0.2cm} $^\dag$References for optical polarization: Wills et al.\ 1992;

~~Berriman et al.\ 1990; --- for no data available

\hspace*{-0.2cm} $^\ddag$$R^{\ast}$ is the K-corrected ratio of the 5 GHz radio to 2500 \AA -band

~~optical flux densities (Stocke et al. 1992); references for

\hspace*{-0.0cm}~~radio fluxes: V{\'e}ron-Cetty \& V{\'e}ron 2001; 
NVSS (Condon et al.\ 
\hspace*{-0.0cm}~~1998); FIRST (Becker et al.\ 1995; Bauer et al.\ 2000)

\end{table}
\begin{table}
\caption{Positions and magnitudes of the comparison stars}
\begin{tabular}{llrrll} \hline
          &       &             &               &             &               \\
IAU Name  &  Star &  RA(2000)   & DEC(2000)     & ~~$B$         &  ~~$R$          \\
          &       &             &               & (mag)       & (mag)         \\ \hline
          &       &             &               &             &               \\
0003$+$158&  S1   & 00 06 08.42 & $+$16 09 54.4 & 16.31       & 15.32         \\
          &  S2   & 00 06 06.20 & $+$16 10 46.3 & 17.17       & 15.71         \\
          &  S3   & 00 06 05.97 & $+$16 12 15.6 & 16.89       & 15.49         \\
0025$+$307&  S1   & 00 28 25.59 & $+$31 03 19.0 & 15.57       & 14.11         \\
          &  S2   & 00 28 15.86 & $+$31 03 09.8 & 15.57       & 13.89         \\
0043$+$039&  S1   & 00 45 39.87 & $+$04 10 02.0 & 16.95       & 15.50         \\
          &  S2   & 00 45 44.87 & $+$04 10 57.9 & 17.34       & 16.09         \\
          &  S3   & 00 45 44.16 & $+$04 13 26.0 & 17.76       & 15.21         \\
0806$+$315&  S1   & 08 09 06.08 & $+$31 22 19.3 & 16.52       & 15.09         \\
          &  S2   & 08 09 18.58 & $+$31 22 20.7 & 16.54       & 15.11         \\
          &  S3   & 08 09 14.89 & $+$31 20 18.7 & 17.65       & 15.92         \\
0824$+$098&  S1   & 08 27 39.18 & $+$09 41 13.5 & 16.60       & 15.03         \\
          &  S2   & 08 27 44.30 & $+$09 45 05.6 & 16.28       & 15.44         \\
0832$+$251&  S1   & 08 35 26.47 & $+$24 57 12.2 & 18.86       & 16.62         \\
          &  S3   & 08 35 47.24 & $+$24 57 19.0 & 16.56       & 15.72         \\
0846$+$513&  S1   & 08 50 14.07 & $+$51 06 21.9 & 17.27       & 16.11         \\
          &  S3   & 08 50 19.88 & $+$51 09 00.0 & 17.34       & 18.67         \\
0850$+$440&  S1   & 08 53 28.75 & $+$43 46 22.8 & 18.37       & 16.08         \\
          &  S2   & 08 53 48.92 & $+$43 48 28.1 & 18.03       & 16.39         \\
          &  S3   & 08 53 39.97 & $+$43 46 15.4 & 18.72       & 16.46         \\
0931$+$437&  S1   & 09 34 46.90 & $+$43 32 05.9 & 15.75       & 14.42         \\
          &  S2   & 09 35 01.19 & $+$43 27 43.4 & 15.72       & 15.24         \\
0935$+$416&  S1   & 09 38 40.37 & $+$41 26 11.3 & 16.11       & 15.32         \\
          &  S3   & 09 39 02.53 & $+$41 30 37.9 & 16.27       & 15.47         \\
0945$+$438&  S1   & 09 49 28.88 & $+$43 37 54.4 & 15.30       & 14.35         \\
          &  S3   & 09 49 06.74 & $+$43 29 08.2 & 17.18       & 15.27         \\
          &  S4   & 09 48 58.30 & $+$43 55 11.8 & 17.28       & 16.14         \\
1029$+$329&  S1   & 10 32 10.68 & $+$32 36 08.1 & 16.37       & 15.02         \\
          &  S2   & 10 32 07.49 & $+$32 37 28.2 & 17.35       & 15.33         \\
1418$+$546&  S1   & 14 20 02.31 & $+$54 25 25.3 & 16.28       & 15.58         \\
          &  S2   & 14 19 46.29 & $+$54 26 43.4 & 16.11       & 15.51         \\
          &  S3   & 14 19 39.75 & $+$54 21 56.1 & 16.74       & 15.44         \\
1422$+$424&  S1   & 14 25 03.56 & $+$42 14 41.8 & 16.27       & 15.67         \\
          &  S2   & 14 25 09.20 & $+$42 17 21.8 & 18.14       & 16.81         \\
          &  S3   & 14 25 11.09 & $+$42 17 51.2 & 15.97       & 15.39         \\
1425$+$267&  S1   & 14 27 47.53 & $+$26 35 14.9 & 15.21       & 13.65         \\
          &  S2   & 14 27 30.07 & $+$26 36 05.5 & 15.59       & 14.01         \\
1444$+$407&  S1   & 14 16 54.96 & $+$40 36 51.9 & 15.68       & 14.10         \\
          &  S2   & 14 46 55.62 & $+$40 36 16.6 & 17.04       & 14.99         \\
1522$+$101&  S1   & 15 24 03.25 & $+$09 58 15.2 & 16.99       & 15.03         \\
          &  S2   & 15 24 07.32 & $+$10 01 02.9 & 17.12       & 15.56         \\
1553$+$113&  S1   & 15 55 35.71 & $+$11 09 33.2 & 16.11       & 15.11         \\
          &  S3   & 15 55 51.81 & $+$11 12 28.7 & 16.55       & 15.45         \\
1631$+$395&  S1   & 16 33 01.57 & $+$39 20 49.4 & 17.20       & 15.90         \\
          &  S3   & 16 32 54.19 & $+$39 21 19.8 & 17.91       & 16.48         \\
1750$+$507&  S1   & 17 51 07.39 & $+$50 45 03.5 & 20.11       & 19.55         \\
          &  S2   & 17 51 06.32 & $+$50 44 33.9 & 16.38       & 14.81         \\
          &  S3   & 17 51 37.59 & $+$50 43 56.5 & 15.70       & 14.80         \\
          &       &             &               &             &               \\ \hline
\end{tabular}

\end{table}

The observations under this program typically achieved convincing
detectability of INOV at a level of 0.01--0.02 mag and spanned a total
of 113 nights (720 hours) between October 1998 and May 2002.
This work provided the first positive detection of INOV for RQQs,
though modest evidence for such variations had been obtained
earlier (e.g.\ Gopal-Krishna et al.\ 1995, 2000; Sagar et al.\ 1996;
Jang \& Miller 1997).
It was, moreover, found that except for BLs and high
optical polarization CDQs (HP-CDQs), the amplitude of detected INOV is
small ($\le$ 3 per cent) and so is the INOV duty cycle ($\sim$10--20 per cent),
{\it irrespective of the radio loudness}. Further, for the BLs and HP-CDQs,
for which a strong INOV was frequently observed, no correlation was
found between the amplitude of INOV and long-term optical
variability (St04b).  We argued that these results are consistent
with the hypothesis that even radio-quiet QSOs possess relativistic
jets emitting optical emission on sub-parsec scales, but that we
are observing them at moderately large angles to the jet direction
so that any variations are neither amplified in magnitude nor
compressed in time-scale as they are in BL Lacs
(GK03; St04a).  Wills (1996) also argued that RQQs do indeed
possess jets, but that they propagate through denser gas close to
the host galaxies' planes and are thus quickly snuffed out.
For BLs we found no correlations between apparent brightness levels and INOV
properties (St04b).  This is in accord with a recent study which
indicates that  microvariability of a blazar may be correlated with
the presence of longer-term flux changes, rather than its apparent
brightness level (Howard et al.\ 2004).

In this paper, we present the results of our
optical monitoring for another 20 AGN belonging to all of the above
mentioned four classes of luminous AGN. We then combine the present data
for 3 BLs with similar high quality light-curves for another 9 BLs taken
from literature, to arrive at a
representative sample of 51 intra-night optical light-curves for
BL Lacs.  This sample allows us to make a  
comparative study of the
INOV properties of BL Lacs detected with the {\it EGRET} instrument on the
{\it Compton Gamma-Ray Observatory} (Hartman et al.\ 1999), or otherwise
found to emit high energy $\gamma$-rays, and their counterparts that
were not detected by {\it EGRET} (Section 4).  Our conclusions are
summarized in Section 5.

\section{Intranight optical monitoring}

All AGNs chosen for this additional study had to be bright
enough to allow a high temporal density for precision differential
photometry using telescopes of a modest aperture. This led to a
requirement that $m_B < 17$ mag. We also wanted to minimize the
contamination problems that arise when the host galaxy contributes
a significant portion of the visible light (e.g.\ Cellone, Romero
\& Combi 2000), and so restricted our sample to 
luminous AGNs (quasars), with $M_B < -23.5$ mag. 
For good visibility from India the sources had to be
at moderate positive declinations and within suitable ranges of
right ascensions.  Basic  data on our sources is presented in Table 1.
Twelve of the sources are RQQs (using the usual criterion for
the K-corrected ratio of 5 GHz to 2500{\AA} fluxes, $R^{\ast}
< 10$), 4 are LDQs, 1 is a CDQ, and 3 are BLs; their redshifts range
from 0.22 to 1.97.

The majority of the data was obtained at ARIES (formerly UPSO), Naini Tal, India, 
using the 104-cm Sampurnanand telescope  
which is an RC system with a f/13 beam 
(Sagar 1999). The detectors used
were a cryogenically cooled  1024 $\times$ 1024 CCD chip (prior to
October 1999) and  a 2048 $\times$ 2048 chip (after October 1999),
both mounted at the cassegrain focus. The 1k$\times$1k chip has a readout 
noise of 7 electrons and a gain of 11.8 electrons/ADU, whereas the 
2k$\times$2k chip has a readout noise of 5 electrons
(in the usually employed slow readout mode) and a gain of 10
electrons/ADU.  Each pixel of both of these CCDs correspond to
0.38 $\times$ 0.38 arcsec$^2$ on the sky, covering a total field of
12$^\prime$ $\times$ 12$^\prime$ in the case of the larger CCD and
6$^\prime$ $\times$ 6$^\prime$ in the case of the smaller CCD
(Sagar 1999).  An R Cousins filter was used for these observations.
On each night only one AGN was monitored, as continuously
as possible. The choice of exposure times depended on the
brightness of the object, the moon's phase, and sky transparency.
The field containing the AGN was adjusted so as to have within the
CCD frame at least 2 (usually 3 or more) comparison stars
within about a magnitude of the AGN;  for nearly all objects 
we were able to find at least one steady comparison star fainter,
or $<$ 0.4 mag brighter, 
than the AGN, so as to obtain an equivalent S/N in the CCD frames.
Seeing ranged from about 1.5 to
about 3.5 arcsec.

Four of the RQQs in the sample were monitored in V Johnson 
passband using a cryogenically cooled Tektronix CCD detector at the
f/3.23 prime focus of 2.34 meter Vainu Bappu Telescope (VBT) of the
Indian Institute of Astrophysics, at Kavalur, India (Table 3). 
The chip has 1024 $\times$ 1024 pixels of approximately 24
$\times$ 24 $\mu$m$^2$, with each pixel dimension corresponding to about
$0.63$ arcsec on the sky, so that the total area covered by a CCD frame
is 10.75$^\prime$ $\times$ 10.75$^\prime$. The readout noise was 4
electrons and the gain was 4 electrons/ADU. Typical seeing was around
2 arcsec.

One night of monitoring data for the RQQ 1422+424 reported here was carried out in V
Johnson passband using the 
Tektronix 1k$\times$1k CCD detector at the f/13 Cassegrain focus of the 1.2
meter Gurushikhar Telescope (GSO) at Mount Abu, India (Table 3).
Each pixel corresponds to 0.32 arcsec in each dimension and the
entire chip covers approximately 5.4$^\prime$ $\times$ 5.4$^\prime$
of sky.  The readout noise was 4 electrons and gain was 10
electrons/ADU.  Typical seeing was $\sim$1.5 arcsec.

At all three telescopes, observations were carried out in 2$\times$2
binned mode, in order to increase S/N;
 bias frames were taken intermittently and
twilight sky flats were taken for processing of the data.  Initial
processing (bias subtraction, flat-fielding and cosmic ray removal)
as well as photometric reductions 
was done in the usual manner employing
standard routines in IRAF\footnote{Image Reduction and Analysis
Facility, distributed by NOAO, operated by AURA, Inc.  under agreement
with the US NSF.} software.  

Instrumental magnitudes of the AGN and the stars in the images
taken at Naini Tal were obtained using the routines available in the
{\it apphot} package in IRAF. For these reductions, a crucial
parameter,  the circular aperture used for the photometry of the
QSO and the comparison stars, varied from night to night.  For each
night an optimum aperture for the photometry was selected by
considering a range of apertures starting from a minimum corresponding
to the median seeing (FWHM) over the night; we chose the aperture that produced
the minimum variance in the star$-$star differential light curve (DLC)
of the steadiest pair of comparison stars. Additional details of the
observation and reduction procedures are presented elsewhere (Stalin 2003;
St04b).

Instrumental magnitudes of the AGN and stars in the image frames
acquired at VBT and GSO were determined by using DAOPHOT II\footnote{Dominion 
Astrophysical Observatory Photometry Software} (Stetson
1987) and employing aperture photometric techniques. 
The best S/N was found for data reduced with a 7.0 pixel radius
and it is thus used for our analysis. 

The positions and the $B$ and $R$ magnitudes (taken from the  USNO-B
catalog\footnote{http://www.nofs.navy.mil/data/fchpix} (Monet et al.\
 2003) for the comparison stars used in our analysis are given
in Table 2.  Note that the magnitudes of the comparison stars taken
from this catalog have uncertainties of up to 0.25 mag, though errors for individual 
objects are not provided.

\begin{figure*}
\vspace*{-0.5cm}
\hspace*{-0.5cm}\psfig{file=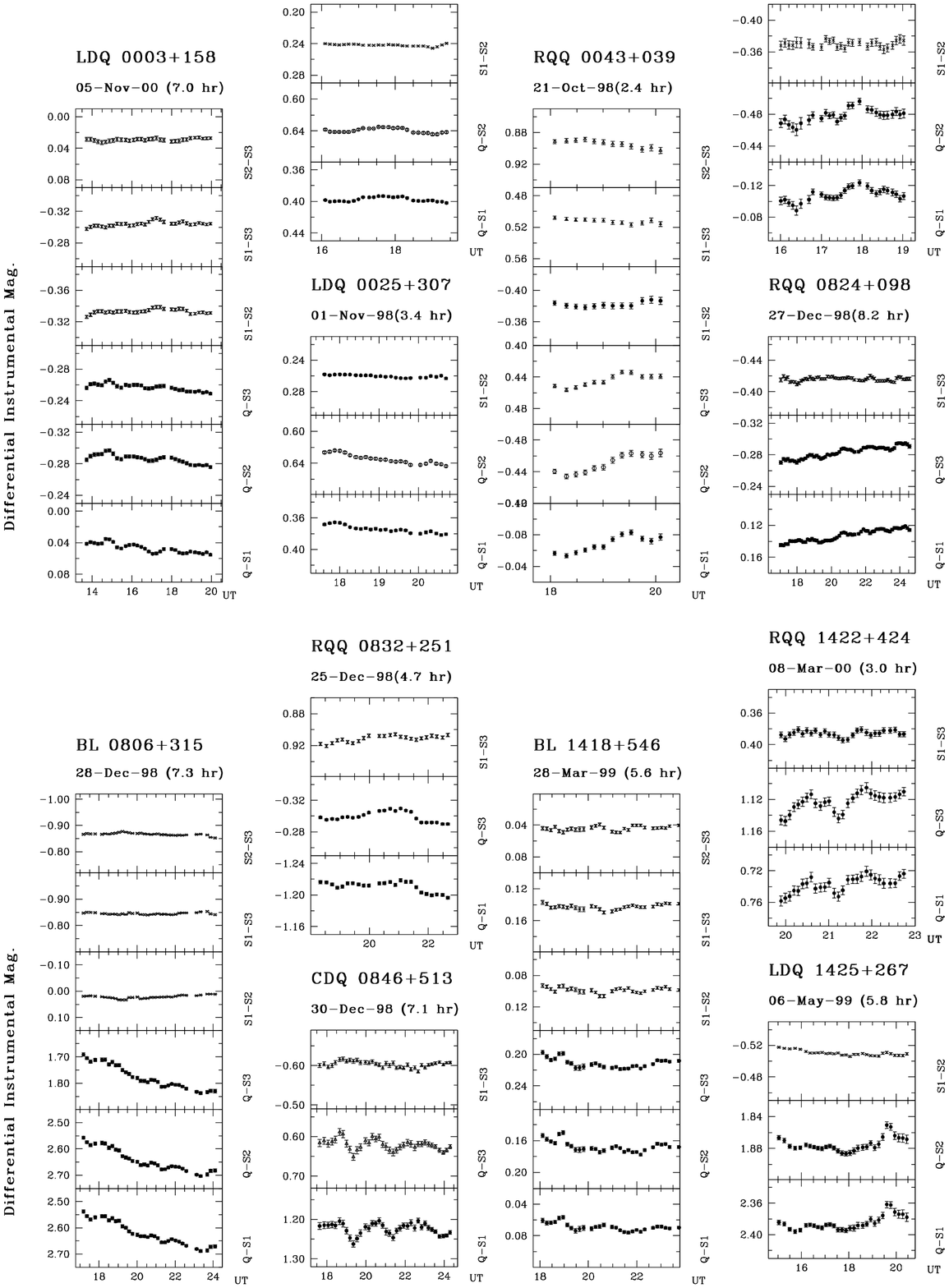,width=18cm,height=23.5cm}
\noindent{\bf Figure 1.} 
 Differential light curves (DLCs) for the quasars on nights
with a positive detection of INOV.
The name of the quasar, the date, and the duration of the observation are given
at the top of each night's data. The upper panel(s) give the 
DLCs for the various pairs of comparison stars
available and the subsequent panels give the quasar-star DLCs, as defined
in the labels on the right side. 
\end{figure*}

\begin{figure*}
\vspace*{-14.5cm}
\hspace*{1.0cm}\psfig{file=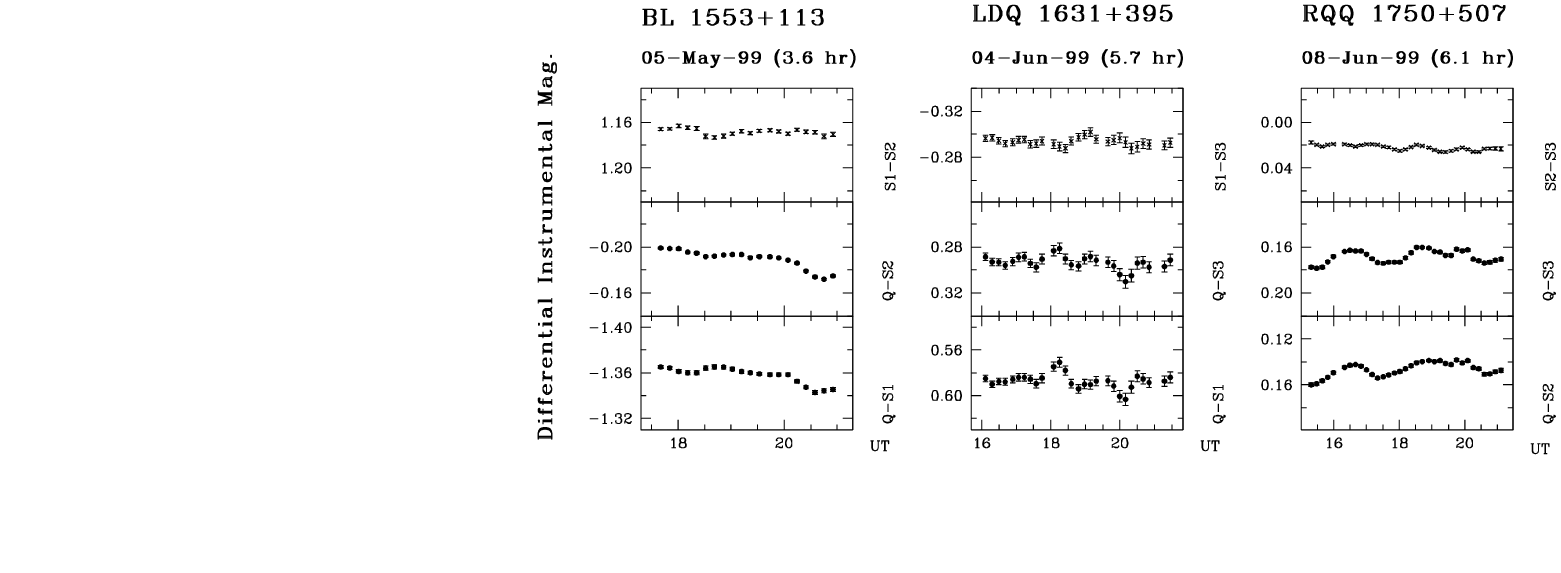}

\vspace*{-8.0cm}
\noindent{\bf Figure 1.} {\it Continued} 
\end{figure*}

\section{Results of intra-night Monitoring}

Figure 1 presents the DLCs for all the nights on which significant
variability was detected for any AGN in the present sample. It can be
seen that in these data, variability of the order of 0.01 mag over the
course of a few hours can be detected. A log of the observations and
the main results are given in Table 3. For each night of observations of
every object, this table
provides the number of data points (Npoints), the duration,
an indicator of the variability status, as well as two
quantitative measures of the variability, $C_{\rm eff}$ and $\psi$ (see below).


\begin{table*}
\caption{Observation log and variability results}
\begin{tabular}{lllccccclrr} \hline
IAU Name  & Other      & Type &  Date     &  Filter& Telescope  & Npoints  & Duration  &  Status$^{\ast}$ & $C_{\rm eff}$ & $\psi$  \\ 
          & Name       &      &           &        &            &          & (hours)   &         &           & (\%)    \\ \hline

0003$+$158& PKS        & LDQ  &  03.11.00 & R      & ARIES       &  28      &  5.8      &  NV     &  ---      &  ---      \\
          &            &      &  05.11.00 & R      & ARIES       &  32      &  7.0      &  V      & 3.1       &  1.8      \\
0025$+$307& RXS        & LDQ  &  13.10.98 & R      & ARIES       &  26      &  3.6      &  V      & 2.7       &  0.8      \\
          &            &      &  01.11.98 & R      & ARIES       &  24      &  3.4      &  V      & 5.1       &  1.9      \\
0043$+$039& PG         & RQQ  &  21.10.98 & R      & ARIES       &  12      &  2.4      &  V      & 4.2       &  2.5      \\
          &            &      &  05.11.98 & R      & ARIES       &  28      &  3.2      &  V      & 2.6       &  3.2      \\
0806$+$315& B2         & BL   &  28.12.98 & R      & ARIES       &  34      &  7.3      &  V      & $>$ 6.6   &  14.5     \\
0824$+$098& 1WGA       & RQQ  &  27.12.98 & R      & ARIES       &  58      &  8.2      &  V      & 4.3       &  2.2      \\
0832$+$251& PG         & RQQ  &  25.12.98 & R      & ARIES       &  24      &  4.7      &  V      & 4.3       &  2.0      \\
          &            &      &  14.01.99 & R      & ARIES       &  63      &  7.3      &  NV     & ---       &  ---      \\ 
          &            &      &  10.12.99 & R      & ARIES       &  31      &  6.7      &  NV     & ---       &  ---      \\ 
0846$+$513& 0846$+$51  & CDQ  &  30.12.98 & R      & ARIES       &  37      &  7.1      &  V      & 2.8       &  5.6      \\
0850$+$440& US 1867    & RQQ  &  17.02.99 & R      & ARIES       &  37      &  7.7      &  NV     & ---       &  ---      \\
0931$+$437& US 737     & RQQ  &  20.02.99 & R      & ARIES       &  24      &  4.5      &  NV     & ---       &  ---      \\
0935$+$416& PG         & RQQ  &  27.03.99 & R      & ARIES       &  15      &  2.7      &  NV     & ---       &  ---      \\
0945$+$438& US 995     & RQQ  &  15.01.99 & V      & VBT        &  10      &  2.2      &  NV     & ---       &  ---      \\
1029$+$329& CSO 50     & RQQ  &  13.03.99 & V      & VBT        &  57      &  5.4      &  NV     & ---       &  ---      \\
1418$+$546& OQ 530     & BL   &  28.03.99 & R      & ARIES       &  31      &  5.6      &  V      & 4.0       &  2.0      \\
1422$+$424& RXS        & RQQ  &  03.04.99 & R      & ARIES       &  39      &  7.2      &  NV     & ---       &  ---      \\
          &            &      &  14.04.99 & V      & VBT        &  40      &  4.1      &  NV     & ---       &  ---      \\ 
          &            &      &  07.03.00 & R      & ARIES       &  15      &  3.9      &  NV     & ---       &  ---      \\  
          &            &      &  08.03.00 & V      & GSO        &  28      &  3.0      &  V      & 2.9       &  3.6      \\ 
1425$+$267& B2         & LDQ  &  06.05.99 & R      & ARIES       &  31      &  5.8      &  V      & 2.8       &  3.2      \\
1444$+$407& PG         & RQQ  &  15.04.99 & V      & VBT        &  28      &  2.9      &  NV     & ---       &  ---      \\ 
1522$+$101& PG         & RQQ  &  11.04.99 & R      & ARIES       &  36      &  6.6      &  NV     & ---       &  ---      \\
1553$+$113& PG         & BL   &  05.05.99 & R      & ARIES       &  20      &  3.6      &  V      & $>$6.6    &  2.3      \\
          &            &      &  06.06.99 & R      & ARIES       &  40      &  7.1      &  NV     & ---       &  ---      \\ 
1631$+$395&KUV         & LDQ  &  04.06.99 & R      & ARIES       &  28      &  5.7      &  V      &  2.9      &  2.7      \\
          &            &      &  30.05.00 & R      & ARIES       &  12      &  3.5      &  NV     &  ---      &  ---      \\ 
1750$+$507& IRAS       & RQQ  &  03.06.98 & R      & ARIES       &  44      &  4.7      &  NV     &  ---      &  ---      \\ 
          &            &      &  06.06.98 & R      & ARIES       &  15      &  1.6      &  NV     &  ---      &  ---      \\ 
          &            &      &  08.06.99 & R      & ARIES       &  34      &  6.1      &  V      &  $>$6.6   &  2.0      \\  \hline
\end{tabular}

\hspace*{-10.7cm}$^{\ast}$V = variable; NV = non-variable

\end{table*}

The parameter $C_{\rm eff}$  is defined,  basically following
Jang \& Miller (1997), for a
given DLC as the ratio of the standard deviation of all its data points, $\sigma_T$, to the
averaged standard deviation for its individual data points, $\sigma = \eta\sigma_{\rm err}$.
Here $\eta$ is the factor by which the average of the measurement 
errors ($\sigma_{\rm err}$, as given by {\it phot}) should be 
multiplied; we find $\eta$ = 1.50 (Stalin 2003; St04b).
We compute $C_{\rm eff}$ from the $C_i$ values (defined as the ratio of the 
standard deviation of $i^{th}$ DLC to the mean $\sigma$ of its 
individual data points multiplied by the factor $\eta$) determined 
for the DLCs of an AGN relative to different comparison stars,
measured on a single night (see Sa04 for details).
A value of $C_{\rm eff} > 2.57$ corresponds to a confidence
level of variability in excess of 0.99 and is the criterion we
use to assign variability to a QSO.  
We note that for these AGN all the DLCs involving only their 
comparison stars were found to show statistically insignificant
variability, using the same statistical criterion.

We quantify the actual variation of the QSO on a given night
using the error corrected amplitude of variability, $\psi$, as defined by
Romero, Cellone \& Combi (1999),
\begin{equation}
\psi = \sqrt{(D_{max} - D_{min})^2 - 2\sigma^2},
\label{amplitude}
\end{equation}
with $D_{max}$ ($D_{min}$)  the maximum (minimum) in the quasar 
DLC, and $\sigma$ the corrected error value described in the previous
paragraph. Details are given in St04b.  

The structure function (SF) is frequently used to characterize 
variability properties such as time-scales and periodicities 
present in the light-curves.
We have also computed the SFs for our dataset
in the fashion discussed in some detail in Sa04 and St04b.
Basically, a monotonically rising SF indicates that the source
shows no temporal structure on time scales shorter than the
duration of the light curve, while the beginning of a plateau in the 
SF signifies a time-scale for the variability and a dip
in the SF may be indicative of a periodic component.
 Figure 2 shows
the SFs for five objects on the nights when
they were rather strongly variable, with $\psi > 0.03$ mag. 

\begin{figure}
\hspace*{0.5cm}\psfig{file=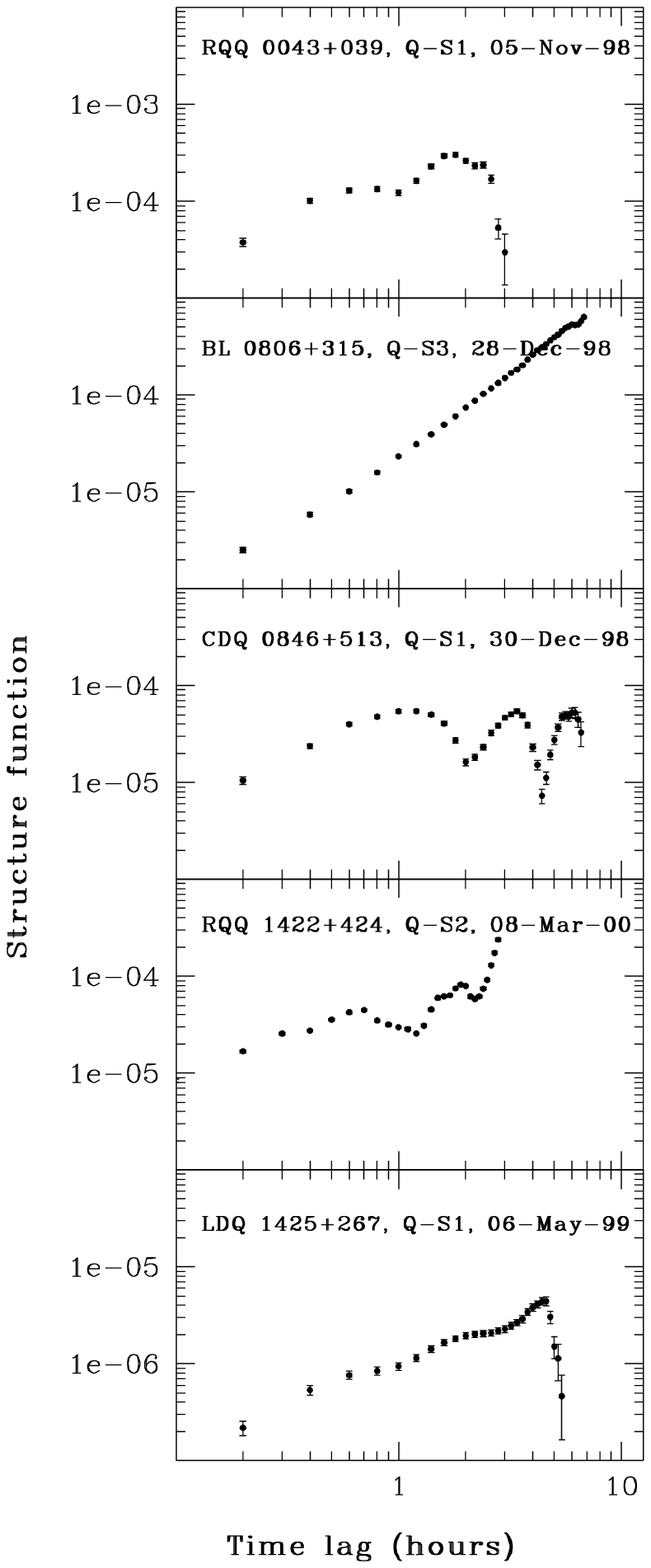,width=8.0cm,height=15cm}
\noindent{\bf Figure 2.} 
Structure functions for the 5 most strongly variable 
quasar light curves.
The object, DLC descriptor and date label each of the panels.
\end{figure}

We now give brief comments on a few of the sources  
that showed INOV.

{\bf RQQ 0043$+$035} 
varied on both the nights it was observed; on the first night it
brightened by $\sim 0.02$ mag in $\sim 1$ hour.  On the second night,
about 2 weeks later, the data were relatively noisy; nonetheless,
a brightening by about $\sim 0.03$ mag over 2 hours is clearly
detected.  The structure function for this night shows a time
scale of roughly 1.5 hours (Fig.\ 2).

{\bf BL 0806$+$315}: 
on the single night this BL Lac was
monitored for about 7 hours, a fading by about 0.15
mag was detected. The SF indicates that no time scale shorter than the
monitoring duration is present (Fig. 2).

{\bf CDQ 0846$+$513} 
is the only core-dominated 
quasar in the present sample. A fluctuation of $\sim$ 0.05 mag
can be seen on its DLC. The SF shows hints of periodicities of about 2 and
4.5 hours (Fig.\ 2); but 
the data train is much too short to justify claiming them as actual
periodicities.  As this is a gravitationally lensed quasar (e.g.
Maoz et al.\ 1993), much of its variability may be extrinsic,
produced by microlensing.

{\bf RQQ 1422$+$424} 
showed variability on just one of
the 4 nights it was monitored. The DLC in Fig.\ 1 shows a quasi-oscillatory
pattern, with an amplitude of $\sim$ 0.04 mag. The SF
suggests a timescale of $\sim$ 1 hour (Fig.\ 2).

{\bf LDQ 1425$+$267} 
showed a weak flare of $\sim$ 0.02
mag near 19.6 UT on the single night it was monitored. 
The SF hints at a time scale of $\sim$ 4 hours (Fig. 2).

In order to obtain more significant estimates of the INOV duty
cycle (DC), we have combined the results for the 12 RQQs,
4 LDQs and 3 BLs in this sample with the extensive monitoring
data presented for these AGN classes in our earlier work
(St04a, Sa04). It may be recalled that for a
given class of objects, the DC is defined as the weighted
fraction of its DLCs which show INOV, where the contribution
of an individual DLC to this fraction is weighted inversely
by the duration of that DLC in the emitter's frame (Romero et al.\ 1999; 
GK03; St04a,b; Sa04). 
Using these enlarged samples based on
our observations, we estimate DCs of 22 per cent, 22 per cent,  and 63 per cent for
RQQs, LDQs, and BLs, respectively. 

\section{Statistics of INOV in BL Lacs using enlarged datasets}
Although INOV of blazars has been clearly established for
about 15 years (Miller et al. 1989; Carini 1990), only recently
has enough data been accumulated to allow a reliable description
of its frequency and amplitude, and to examine if various AGN
classes exhibit different INOV behaviour. In order to increase
the sample of light curves (LCs) we have combined the results
for BLs in this paper with those reported in Sa04 and in
other papers taken from the literature from 1990 through 2003
reporting intra-night optical monitoring of BLs (as classified
in the V\'eron-Cetty \& V\'eron catalogue, 2001). 
Note that the object 0537$-$441 may also be classified as a CDQ (see
Maraschi et al.\ 1985); however, we have considered it to be a BL in
our analysis. 
While it is possible that our literature
search is less than complete, we believe that our selection of
BL Lac monitoring data is both 
extensive and representative. 
We have not included in our sample data from papers
where results are presented for just a
single BL Lac object, nor where the duration of the LC is shorter than 4 hours.
These criteria led to the selection of 51 LCs with durations ranging
between 4 and 10 hours (median = 6.5 hours). These LCs correspond to 12 BLs,
(Table 4) reported in 4 papers: present work (3 LCs, R-filter);
Sagar et al.\  2004 (26 LCs, R-filter); Romero et al.\ 2002 (19 LCs,
V-filter);  Ghosh et al.\ 2001 (3 LCs, B-filter). For the present
purpose, we do not distinguish between data taken using the different
filters. It may be noted that the rms error of individual data points
is typically $\sim$0.003 mag for all the 51 LCs considered.

In Fig. 3 we present the distributions of the INOV amplitude, $\psi$,
for two subsets of the 51 (high quality) LCs. These subsets are derived by
applying the criterion whether or not the LC refers to a BL detected in
$\gamma$--rays with EGRET (Hartman et al.\ 1999), and/or at TeV energies
(Chadwick et al.\ 1999 ).
Henceforth, such BLs will be referred to by the common name EGRET-BLs.
Likewise, BLs not detected at $\gamma$-ray energies will be called non-EGRET-BLs.
A Kolmogorov--Smirnov (K-S) test performed on the two $\psi$ distributions rejects the null
hypothesis that the two distributions are identical; its probability in only 0.038. 
Thus, EGRET BL Lacs
appear to show stronger INOV as compared to 
non-EGRET BL Lacs, though both the number of nights of
observations per object and the total number of objects
is too small to allow this to be a firm conclusion
at this stage. 
If confirmed using larger samples, this
would suggest a stronger Doppler beaming for EGRET BL Lacs.
Possible physical scenarios for this difference 
are mentioned in Sect.\ 5.

\begin{table}
\caption{Consolidated list of the BL Lacs in the extended sample}
\begin{tabular}{lllrrcc} \hline

IAU name  &~$m_B$   &~M$_B$ &$z$~~  &  $P_{opt}^\ast$ &log R$^{\ast}$$^{\dag}$~ & EGRET$^\ddag$\\ 
            &          &         &         & (\%)  &      &       \\ \hline
            &          &         &         &       &      &       \\
	    0219$+$428  &  15.71   & $-$26.5 &    0.444 &11.7  & 2.8  & Yes \\
	    0235$+$164  &  16.46   & $-$27.6 &    0.940 &14.9  & 3.4  & Yes \\
	    0414$+$009  &  16.86   & $-24.6$ &    0.278 & 2.8  & 2.2  & No  \\
	    0537$-$441  &  17.00   & $-$27.0 &    0.894 &10.5  & 3.8  & Yes \\
	    0735$+$178  &  16.76   & $-$25.4 & $>$0.424 &14.1  & 3.5  & Yes \\
	    0806$+$315  &  15.70   & $-$25.0 &    0.220 & ---  & 1.7  & No  \\
	    0851$+$202  &  15.91   & $-$25.5 &    0.306 & 12.5 & 3.3  & Yes \\
	    1215$+$303  &  16.07   & $-$24.8 &    0.237 &8.0   & 2.6  & No  \\
	    1308$+$326  &  15.61   & $-$28.6 &    0.997 &10.2  & 2.8  & No  \\
	    1418$+$546  &  16.17   & $-$23.7 &    0.151 & 7.5  & 3.1  & No  \\
	    1553$+$113  &  15.00   & $-$26.8 &    0.360 & ---  & 2.2  & No  \\
	    2155$-$304  &  13.36   & $-$25.9 &    0.116 & 4.9  & 1.5  & TeV \\
\hline
\end{tabular}
\hspace*{-0.0cm} $^\ast$References for optical polarizations: Wills et al.\ 1992; Impey \& 

~~~Tapia 1988; Marcha et al.\ 1996; --- implies no data available

\hspace*{0.0cm} $^\dag$$R^{\ast}$ is the K-corrected ratio of the 5 GHz radio to 2500 \AA -band 

~~~optical flux densities (Stocke et al. 1992); reference for radio 

~~~fluxes: V{\'e}ron-Cetty \& V{\'e}ron 2001

\hspace*{-0.0cm} $\ddag$Reference for EGRET detections:  Hartman et al.\ 1999; 

~~~for TeV detection: Chadwick et al.\ 1999 

\end{table}

\begin{figure}
\psfig{file=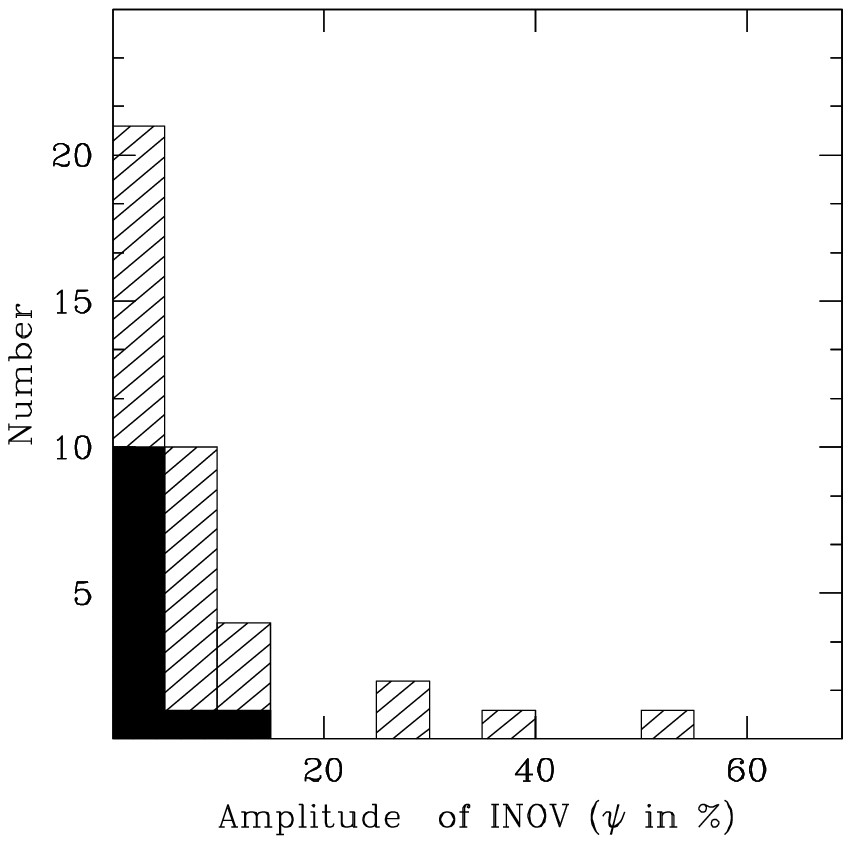}
\noindent{\bf Figure 3.} Distributions of INOV amplitude
($\psi$), for 39 light-curves of the 6 EGRET (hatched) and 12 light-curves of the 6 non-EGRET (black) BLs. 

\end{figure}

\section{Conclusions}

We have presented new observations of intra-night optical
monitoring for 20 powerful AGN, including 3 BL Lacs, 5
radio-loud quasars (RLQs) and 12 radio-quiet quasars (RQQs).
INOV is detected in all three classes of AGN, consistent
with the results reported in our earlier papers (GK03; St04a,b; Sa04).
By combining the present data with the observations reported
in our earlier papers (GK03; St04a; Sa04), we could assemble a
larger AGN sample consisting of 19 RQQs, 9 BL Lacs and 16 RLQs (after 
excluding the high optical polarization quasar 1216$-$010
from the radio core-dominated RLQs in our sample).

The INOV duty cycles (DCs) derived for this sample are:
63\% for BL Lacs, 18\% for RLQs and 22\% for RQQs. Thus,
the INOV duty cycles for both RQQs and RLQs (5 of which
are CDQs) are similar, and much smaller
than that for BL Lacs. This supports our earlier result that
the mere presence of a powerful radio synchrotron jet does
not lead to an enhanced INOV (GK03; St04a). 
The observed similarity in the INOV of RQQs and non-blazar RLQs, 
both in terms of DC and $\psi$, further 
suggests that the RQQs also eject relativistic
jets.  Their jets are, however, probably quenched while crossing
the innermost micro-arcsecond scale, possibly through heavy
inverse Compton losses in the vicinity of the central engine
(GK03). A similar conclusion has also been reported recently from radio 
variability studies of RLQs and RQQs (Barvainis et al.\ 2004).

Further, we have formed an enlarged sample of BL Lac
objects with intra-night monitoring duration $>$ 4 hours, by combining 
the 3 LCs reported here with 48 taken from the literature (Sect.\ 4).
The duty cycle of INOV for this entire sample
of 51 LCs of BL Lacs is found to be 68\%. 

Dividing this sample of 51 LCs by the criterion of
detection of $\gamma-$rays (Table 4), we find that the 
$\gamma-$ray detected BLs show somewhat stronger INOV, the formal
confidence being 0.962 using K-S test (Fig.\ 3). It is
tempting to speculate about the possible origin of this
difference. The synchrotron self-Compton (SSC) model for
the origin of $\gamma-$rays posits that the $\gamma-$rays
are produced by inverse Compton (IC) scattering of
the synchrotron photons themselves off the
relativistic jet electrons (e.g.\ Maraschi, Ghisellini \&
Celloti 1992; Bloom \& Marscher 1996).  The external Compton
(EC) models invoke IC scattering of 
photons originating
outside the jet, typically from the accretion disk around the
central black hole (e.g.\ Dermer, Schlickeiser \&
Mastichiadis 1992), or disk photons
reprocessed by matter above the disk but outside the jet (e.g.
Sikora, Begelman \& Rees 1994; Blandford \& Levinson 1995).
A variant of the EC model, the ``mirror model'', utilizes jet photons
reflected or reprocessed by clouds external to the
jet (Ghisellini \& Madau 1996).   

In the IC scenario involving external seed photons for $\gamma$--ray
loud blazars, one expects the emission cone to be particularly sharp (Dermer
1995), raising the likelihood of detecting stronger and more rapid 
INOV (GK03). Hence, our current preliminary results 
provide additional support to the EC
model. Nonetheless, more extensive, multi-band intra-night monitoring
observations of blazars are clearly needed. For the non-blazar AGN,
the amplitude of INOV continues to be found to be small ($<$ 3\%),
emphasizing the need for even more sensitive monitoring programs.

\section*{Acknowledgements}
It is a pleasure to thank Amy Campbell, Dick Miller and
Angela Osterman for helpful conversations, and the referee
for useful suggestions. CSS is grateful
for  hospitality at  Georgia State University. GK
would like to thank the Institute d'Astrophysique (Paris)
for hospitality during this work. PJW is grateful for
hospitality at Princeton University Observatory while
this paper was written. PJW's efforts are partially
supported by RPE funds to PEGA at GSU.

\end{document}